\documentclass[showpacs,preprintnumbers,amsmath,amssymb]{revtex4}
\usepackage{epsfig}
\usepackage{axodraw}
\usepackage{graphicx}% Include figure files
\usepackage{dcolumn}% Align table columns on decimal point
\usepackage{bm}% bold math
\newcommand{\be}{\begin{equation}}
\newcommand{\ee}{\end{equation}}
\newcommand{\ba}{\begin{eqnarray}}
\newcommand{\ea}{\end{eqnarray}}
\newcommand{\jj}{J/\psi}
\begin{document}
%\voffset=1truecm
\preprint{ROME1-1369/2004, FNT/T 2004/1, LAPTH-1027/04, CERN-PH-TH/2004-037}

\title{$J/\psi$ Absorption in  Heavy-Ion Collisions}% Force line breaks with \\
\author{L. Maiani}
\email{luciano.maiani@roma1.infn.it}
\affiliation{Dipartimento di Fisica, Universit\`{a} di Roma ``La Sapienza''}
\affiliation{Istituto Nazionale di Fisica Nucleare, Sezione di Roma, 
I-00185 Roma, Italy}
\author{F. Piccinini}
\email{fulvio.piccinini@pv.infn.it}
\affiliation{Istituto Nazionale di Fisica Nucleare, Sezione di Pavia}
\affiliation{Dipartimento di Fisica Nucleare e Teorica, Via A. Bassi 6, I-27100,
Pavia, Italy} 
\author{A.D. Polosa}
\email{antonio.polosa@cern.ch}
\affiliation{LAPTH, 9, Chemin de Bellevue, BP 110, 74941 Annecy-le-Vieux Cedex, France}
\author{V. Riquer}
\email{veronica.riquer@cern.ch}
\affiliation{CERN, Department of Physics, Theory Division, 
Geneva, Switzerland}

\date{\today}% It is always \today, today,

\begin{abstract}
We present a new calculation of the
$\pi$--$J/\psi$ dissociation cross sections within 
the Constituent Quark--Meson Model recently introduced. 
To discuss the absorption of $J/\psi$ in heavy-ion 
collisions, we assume the  $J/\psi$ to be 
produced inside a thermalized pion gas, as discussed by Bjorken, 
and introduce the corrections due to absorption by nuclear 
matter as well. We fit the absorption length of the 
$J/\psi$ to the data obtained at  the CERN SPS by the NA50 Collaboration for 
Pb-Pb collisions.
Collisions of lower centrality allow us to determine the temperature and 
the energy density of the pion gas. For both these quantities we find 
values close to those indicated by lattice gauge 
calculations for the transition to a quark--gluon plasma. A simple 
extrapolation to more central collisions, which takes into account the 
increase of the energy deposited due to the increased nucleon flux, 
fails to reproduce the break in  $J/\psi$ absorption 
indicated by NA50, thus lending support to the idea that an unconfined 
quark--gluon phase may have been produced. This conclusion could be 
sharpened by analysing  in a similar way, as a function of centrality,  
other observables such as strange particle production.

\pacs{25.75.-q, 12.39.-x}
\end{abstract}

\maketitle

\section{Introduction} 

In a recent paper~\cite{couplings} the couplings of $J/\psi$ 
to 
$D^{(*)}$ mesons (where by $D^{(*)}$ we mean $D$ or $D^*$)
and to $\pi$'s have been computed in the framework of a 
Constituent Quark--Meson Model introduced in~\cite{proto}. 
The interest of this study stems from the possibility 
that $J/ \psi$ absorption processes of the type: 
\begin{equation} 
\pi +J/\psi\to D^{(*)}+ \bar{D}^{(*)}\label{1}\;\;,
\end{equation} 
play an important role in the relativistic heavy-ion scattering. Since a decrease in 
the $J/\psi$ production in these processes might signal the formation of Quark--Gluon 
Plasma (QGP), it is useful to have reliable estimates of the cross sections 
for the processes (\ref{1}), which provide an alternative way to reduce the $J/\psi$ 
production rate. Previous studies of these effects can be found in~\cite{cina}. 
In this paper we will address the problem of understanding how far we can go
in explaining $\jj$ suppression data via an hadronic mechanism such as (\ref{1}).
Does the hot gas of pions formed after the heavy-ion collision provide a
source of attenuation of $\jj$ antagonist to the standard QGP suppression?

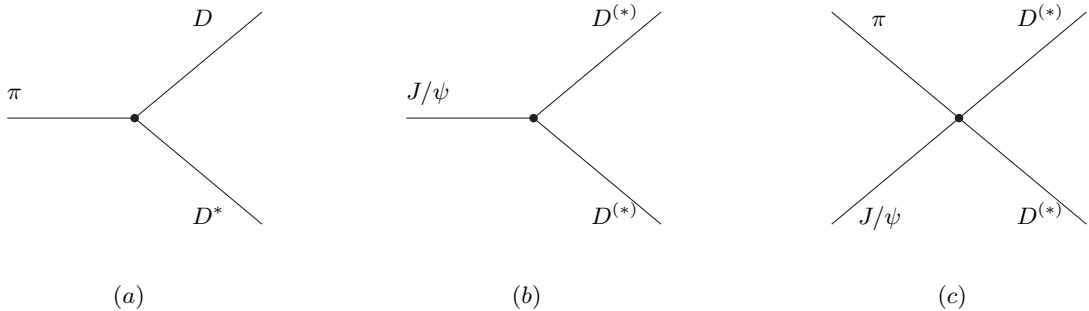
\begin{figure}[bth]
\SetScale{0.8}
\noindent
{\begin{picture}(300,100)(0,0)
\Line(120,0)(60,50)
\Line(60,50)(0,50)
\Vertex(60,50){2}
\Line(120,100)(60,50)
\put(0,47.5){$\pi$}
\put(70,75){$D$}
\put(70,0){$D^*$}
\put(40,-30){$(a)$}
\end{picture}
\hspace*{7em}
\begin{picture}(0,0)(220,0)
\Line(120,0)(60,50)
\Line(60,50)(0,50)
\Vertex(60,50){2}
\Line(120,100)(60,50)
\put(0,47.5){$\jj$}
\put(70,75){$D^{(*)}$}
\put(70,0){$D^{(*)}$}
\put(40,-30){$(b)$}
\end{picture}
\hspace*{7em}
\begin{picture}(0,0)(130,0)
\Line(120,0)(60,50)
\Line(60,50)(0,0)
\Line(60,50)(0,100)
\Vertex(60,50){2}
\Line(120,100)(60,50)
\put(15,75){$\pi$}
\put(70,75){$D^{(*)}$}
\put(70,0){$D^{(*)}$}
\put(10,0){$J/\psi$}
\put(40,-30){$(c)$}
\end{picture}
}
\vskip 1truecm
\caption{\label{fig.0} \footnotesize  Couplings involved in the
tree--level calculations of processes (\ref{1}).}
\end{figure}

The relevant couplings needed to compute the 
cross section of process (\ref{1}) are shown in Fig.~\ref{fig.0}.
%\vskip -0.5truecm
To compute the amplitudes~(\ref{1}),
besides the ${DD^*\pi}$ coupling, Fig.\ref{fig.0}a, whose strength 
$g_{D^*D\pi}$ has been  theoretically estimated and experimentally investigated
~\cite{exp}, one would need also the 
$JD^{(*)}D^{(*)}$, Fig.\ref{fig.0}b, and the $J\,D^{(*)}D^{(*)}\pi$ couplings, 
Fig.\ref{fig.0}c. In an effective Lagrangian approach the latter couplings 
provide direct four-body amplitudes, while the former enter the amplitude {\it 
via} 
tree diagrams with an exchange of a charmed particle $D^{(*)}$ in the 
$t$--channel. 

These couplings have also been estimated by different methods, all presenting, in 
our opinion, different shortcomings. For example the use of the $SU(4)$ symmetry 
puts on the same footing the heavy quark $c$ and the light quarks, which is at odds 
with the results obtained within the Heavy Quark Effective Theory, where the 
opposite approximation $m_c\gg \Lambda_{QCD}$ is used. Similarly, the rather common 
approach based on Vector Meson Dominance should be considered 
critically, given the large 
extrapolation from $p^2=0$ to $p^2=m_{J/\psi}^2$ that 
is involved. A different evaluation, based on QCD Sum Rules can be found in 
\cite{qcdsr} and presents the typical theoretical uncertainties of this method. 
It is useful to have diverse, although inevitably 
model-dependent, calculations to assess a reliable range of theoretical results.
A comparison of our results with the cross sections obtained by other methods 
is given below, so to provide a further indication of the theoretical uncertainties 
(the theoretical uncertainties intrinsic to the model used
in this paper have been discussed in~\cite{couplings}).

To apply our results to  $J/\psi$ production in heavy-ion collisions, we follow the 
picture introduced by Bjorken some time ago~\cite{heavyreaction}.
Seen in the centre-of-mass frame, the colliding nuclei appear as Lorentz-contracted 
pancakes which traverse one another in a short time $t$ (say, $2$~fm/c). 
After crossing, 
the nuclei leave behind them a {\it fireball} with 
essentially vanishing baryon 
number, where the produced $J/ \psi$ is sitting (particles in the 
fireball are often referred to as {\it comovers}).

For low-centrality collisions, the fireball should be, to a good approximation, a 
pion gas that quickly goes to thermal equilibrium with a temperature corresponding 
to the energy deposited by the colliding nucleons in the central region. The energy 
density estimated by Bjorken is in the range of  
$0.3$--$3~{\rm GeV}/{\rm fm}^{3}$ and is 
proportional to the number of nucleons per unit area that participate in the 
collision. 

Our cross section calculation
allows us to compute the absorption length of the $J/ \psi$ in the 
pion gas, which we tentatively approximate with a perfect gas with vanishing 
chemical potential (at high temperature, say $T>150$~MeV, 
there are enough inelastic collisions in the pion gas to provide
chemical equilibrium as well).

To obtain a more accurate value of $T$, we introduce also the 
corrections due to $J/\psi$ absorption by nuclear matter, 
using the absorption cross sections obtained recently from $p-A$ collisions \cite{sigmaN}.

We have fitted the absorption 
length to the data obtained by the   
NA50~\cite{PLNA5099}  collaboration 
at the SPS, 
which show a clear exponential behaviour as a function of the linear dimension of the 
collision region, $l$, up to $l\simeq 5$~fm. 
We find not unreasonable 
values for the temperature ($T=225$~MeV) and for the energy density 
($\epsilon \simeq 0.32~{\rm GeV}/{\rm fm^{3}}$). 
The absorption length decreases very steeply with increasing temperature. 

The energy density is perhaps on the low side with respect to {\it a priori} 
expectations based on the Bjorken formula. On the other hand, the temperature is 
somewhat higher than the critical temperature computed in lattice calculations 
(see~\cite{Karsch} for a recent review). The difference, if any, 
could be due to either an underestimate of the cross section, 
or to the perfect gas approximation being too crude, or 
to both. 
Large departures from perfect gas behaviour are found in numerical simulations below the critical 
temperature~\cite{Karsch}. Further work to improve on the perfect Bose 
gas approximation and obtain a better calibration of the crucial parameters of the 
fireball, temperature and energy density, is needed. 

To extrapolate to higher centrality, we keep into account the increase of the 
average nucleon number per unit area for decreasing impact parameter. This, in 
turn, makes the temperature and the deposited energy increase while the 
absorption length decreases, leading to a downward bending of the $J/ \psi$ 
production as a function of $l$. The 
behaviour we find, however, is too smooth to reproduce the rather sharp break shown 
by the NA50 data for $l$ larger than $5$~fm. This suggests that the simple 
pion gas description ceases to be valid at these values of $l$, thus lending support 
to the formation of a new phase. Our analysis implies that it would be useful to analyse 
in the same way, i.e. as a function of $l$, the data obtained for other observables 
such as the production of strange particles. Given the rather low temperatures that we find 
in the low-centrality region, we would expect very few strange particles to be produced 
in the central region as well. This would not apply to the unconfined phase, where 
the Boltzmann suppression of strange quarks with $m_s = 150$~MeV would not be operative 
for temperatures around $225$~MeV.

\section{Cross Sections}
The calculation of the total cross sections $\sigma(\pi\jj\to D^{(*)}D^{(*)})$
proceeds through the evaluation of the amplitudes listed in the Appendix
and the two-body phase-space integration:
\begin{equation}
\sigma_{\pi\jj\to D\bar{D}}(s)=\frac{M_{\jj}}{12\pi (s-M_{\jj}^2)^2}
\int_R dp_D \frac{p_D}{E_D}\sum_{\rm pol}
|A(s,p_D)|^2\nonumber\;\;,
\end{equation}
where $s$ is the Mandelstam variable $s=(p_\pi + p_{\jj} )^2$, and $R$
is the phase-space interval $R=(p_{D}^{\rm min},p_{D}^{\rm max})$.
An overall factor of 2 is included to count all isospin
quantum numbers in the final state.
The laboratory frame is considered with high-momentum massless 
pions colliding on $\jj$ at rest.  
En energy threshold of about $E_\pi=800$~MeV is required to open the reaction channel.

\begin{figure}[ht]
\begin{center}
\epsfig{%bbllx=0.5cm,bblly=16cm,bburx=20cm,bbury=23cm,
height=7.truecm, width=9.truecm,
        figure=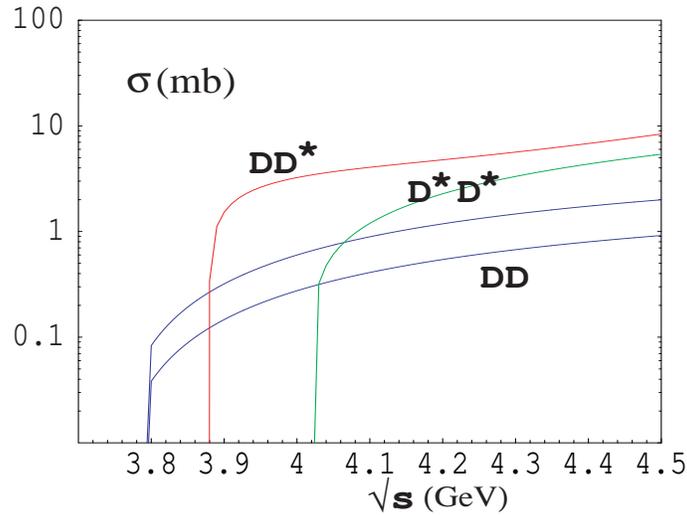}%\vskip1cm
\caption{\label{fig.1} \footnotesize Cross sections 
$\sigma_{\pi\jj}\to (DD,DD^*,D^*D^*)$ as functions of $\sqrt{s}$.
We show the error band for the $DD$ final state resulting from the 
theoretical uncertainty in the determination of the 
couplings~\cite{couplings}. The error bands on $DD^*$ and $D^*D^*$
are similar.}
\end{center}
\end{figure}

For example, in the case of the $D\bar{D}$ final state, using the notation 
introduced in~\cite{couplings}, we have 
$A=A_{1a}+A_{2a}+A_{3a}$, where:
\ba
A_{1a}&=&i\frac{g_{\pi DD^*}g_{\jj DD^*}}{t-M_{D^*}^2} 
\epsilon(\eta,p_4,p_3-p_2,\alpha)\Pi^{\alpha\nu}(p_3-p_2)(p_2)_\nu\nonumber\\
A_{1b}&=&i\frac{g_{\pi DD^*}g_{\jj DD^*}}{u-M_{D^*}^2} 
\epsilon(\eta,p_3,p_1-p_3,\alpha)\Pi^{\alpha\nu}(p_3-p_1)(p_2)_\nu\nonumber\\
A_{1c}&=&i\frac{g_{JDD\pi}}{M_D} \epsilon(\eta,p_2,p_4,p_3)\nonumber,
\ea
where $\eta$ is the polarization vector of $\jj$, $\Pi^{\mu\nu}(q)$ is the
sum-over-polarization-tensor
\begin{equation}
\Pi^{\mu\nu}(q)=\frac{q^\mu q^\nu}{M^2}-g^{\mu\nu},
\end{equation}
$M$ being the mass of the vector particle. The notation 
$\epsilon(\mu,\nu,p,q)= \epsilon_{\mu\nu\rho\sigma}p^\rho q^\sigma$ is used.
The coupling constants $g_{\pi DD^*},g_{\jj DD^*},g_{JDD\pi}$ have been 
discussed in~\cite{couplings} (the notation $g_{JDD\pi}=g_0$ is used), and are:
\ba
g_{\pi DD^*} &=& \frac{13}{1+\frac{q_\pi\cdot p_{\jj}}{\Lambda_1^2}}\nonumber\\
g_{\jj DD^*} &=& 4.05\;\;({{\rm GeV}}^{-1})\nonumber\\
g_{JDD\pi} &=& \frac{234}{1+\frac{q_\pi\cdot p_{\jj}}{\Lambda_2^2}}\;\;({{\rm GeV}}^{-2})\nonumber.
\ea
where $\Lambda_1=1.1$~GeV and  $\Lambda_2=0.8$~GeV. 
Attempts to compute quadrilinear couplings using $SU(4)$
symmetry can be found in~\cite{cina}. The dependence of the 
cross sections on $\sqrt{s}$ for the  $DD$ and $DD^*$ final states are
shown in Fig.~\ref{fig.1}.

\begin{figure}[ht]
\begin{center}
\epsfig{%bbllx=0.5cm,bblly=16cm,bburx=20cm,bbury=23cm,
height=7.truecm, width=8.truecm,
        figure=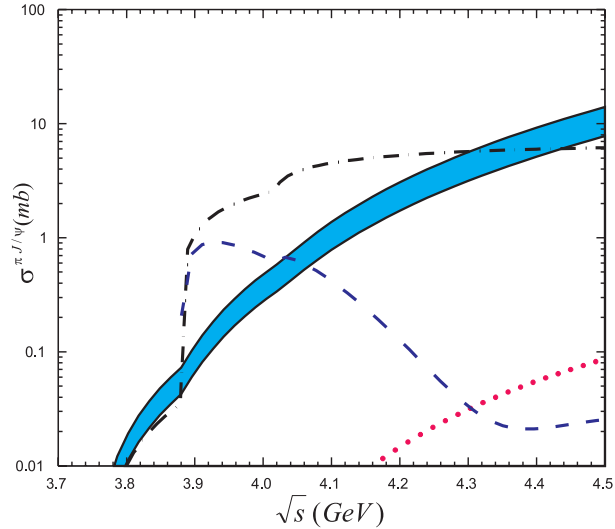}%\vskip1cm
\caption{\label{fig.2} \footnotesize Cross sections
computed with various approaches: QCD sum rules (band), 
short-distance QCD (dotted line), meson-exchange models (dot-dashed line),
non-relativistic constituent quark model (dashed line)~\cite{rapp}.}
\end{center}
\end{figure}

The amplitudes for $DD^*,D^*D^*$ final states are given 
in the Appendix.
The results we obtain can be compared with those shown in Fig.~\ref{fig.2}
extracted by the authors of~\cite{rapp}. Note the sharp rise of $DD^*$ and 
$D^*D^*$, which is due to $S$-wave production.

\section{Thermal Average}
In a first instance we neglect $\jj$ absorption 
by the nuclear matter, and focus on the effect 
of the fireball left over after the nuclei have 
passed each other (i.e. the {\it comoving particles}
in the frame of the target nucleus). The quantity we are interested 
in is the $\jj$ suppression ${\cal A}$ as a function of the linear 
size $l$ of the fireball.
A $\jj$ produced in any internal point of a fireball of diameter $l$ has to 
travel on average a distance $\simeq 6/10 l$ to reach the boundaries of the ball. 
On the other hand the mean free path of the $\jj$ is $\lambda\simeq 1/\rho\sigma$, 
so the attenuation can be defined as:

\begin{equation}
{\cal A}(x) = {\rm exp}\left[-x\langle\rho \cdot \sigma_{\pi\jj\to D^{(*)}D^{(*)}}
\rangle_T\right]\label{eq.attenuation}\;\;,
\end{equation} with $x\simeq 6/10 l$. 

As a simple approximation, the thermal average $\langle... \rangle_T$ is taken in 
a perfect Bose gas of charged and neutral pions with vanishing chemical 
potential:

\begin{equation}
\langle \rho \cdot \sigma_{\pi\jj\to D^{(*)}D^{(*)}}\rangle_T= 
\frac{3}{2\pi^2}\int_{E_\pi^{{\rm th.}}}^{\infty} dE_\pi 
\frac{E_\pi^2 \sigma(E_\pi)}{e^{E_\pi/kT}-1}\;\;.
\end{equation}

For reference, we give also the number density of the relevant pions at temperature 
$T$

\begin{equation}
\rho(T)=\frac{3}{2\pi^2}\int_{E_\pi^{{\rm th.}}}^{\infty} dE_\pi 
\frac{E_\pi^2}{e^{E_\pi/kT}-1}\;\;,
\end{equation} where $E_\pi^{{\rm th.}}$ is the threshold energy required to open 
the reaction channel, and the total energy density: 

\begin{equation}
\epsilon(T)=\frac{3}{2\pi^2}\int_{m_\pi}^{\infty} dE_\pi 
\frac{E_\pi^3}{e^{E_\pi/kT}-1}\;.
\end{equation}

We report in Table I  
the values of $\rho(T)$, $\epsilon(T)$ and $\langle\rho \cdot 
\sigma\rangle_T$ for different values of the temperature. 

\begin{table}[htdp]
\begin{center}
\begin{tabular}{|c|c|c|c|}
\hline
$T$~(MeV) & $\rho$~(fm$^{-3}$) & 
$\epsilon (T)$~(MeV/fm$^3$) & $\langle \rho\sigma\rangle_T$~(fm)$^{-1}$\\
\hline
$205$&$0.0467$&$215 $&$0.037$\\
$215$&$0.0629$&$261 $&$0.051$\\
$225$&$0.0828$&$313 $&$0.070$\\
$235$&$0.11$&$ 373$&$0.093$\\
$245$&$0.13$&$ 441$&$0.12$\\
$255$&$0.17$&$ 518$&$0.16$\\
$265$&$0.20$&$ 605$&$0.20$\\
$275$&$0.25$&$702$&$0.25$\\
\hline
\hline
\end{tabular}
\caption{Values of $\rho(T)$, $\epsilon(T)$ and $\langle\rho \cdot 
\sigma\rangle_T$ for different values of the temperature.} 
\end{center}
\end{table}

\begin{figure}[ht]
\begin{center} \epsfig{%bbllx=0.5cm,bblly=16cm,bburx=20cm,bbury=23cm, 
height=5.truecm, width=8.truecm,  figure=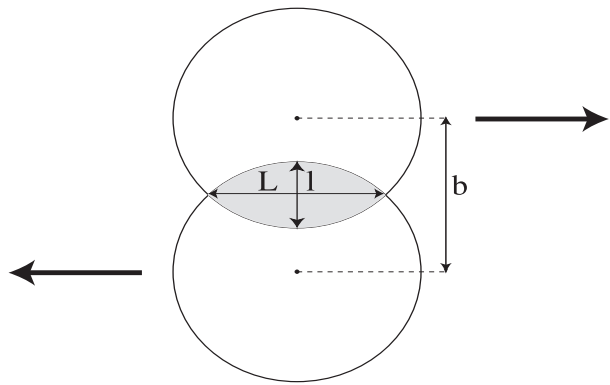}%\vskip1cm 
\caption{\label{spheres} \footnotesize The overlap region
of two colliding equal nuclei. Dimensions $l$ and $L$ determine
the transverse size of the pion fireball and the longitudinal
size of the nucleon column to be traversed by the $\jj$, respectively.
} 
\end{center} 
\end{figure}

To connect to experimental data, we observe that, 
in the approximation of sharp-edged nuclei, 
the linear size of the fireball, $l$,  is given by 
the transverse dimension of the region 
where nuclei overlap (see Fig.~\ref{spheres}), 
which in turn is related to the impact parameter $b$ according to:
\begin{equation}
l= 2R - b\;\;.
\label{eq.Ldef}
\end{equation}

\begin{figure}[ht] 
\begin{center} 
\epsfig{%bbllx=0.5cm,bblly=16cm,bburx=20cm,bbury=23cm, 
height=6.truecm, width=8.truecm, 
        figure=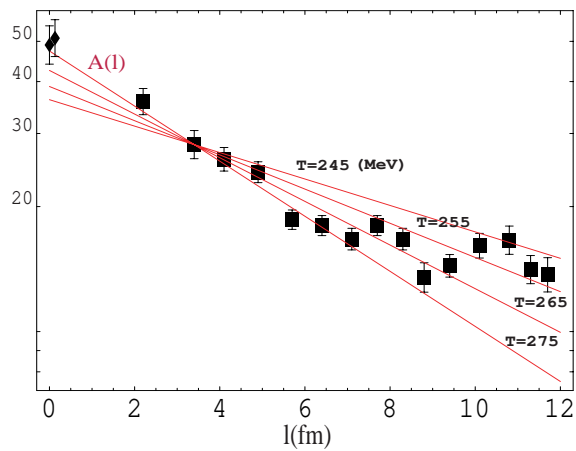}
%attenuation6.eps}%\vskip1cm 
\caption{\label{isotherm} The attenuation function ${\cal A}$ 
given in Eq.~(\ref{eq.attenuation}) 
as a function the linear size of the fireball. 
Lines correspond to a fixed temperature of the $\pi$ gas 
expressed in MeV (isotherms). 
Experimental data: diamonds$=p$--$(p,d)$~(NA51), boxes=Pb-Pb~(NA50).
The value of $l$ for each experimental point is obtained from 
the measured value of the impact parameter $b$ via Eq.~(\ref{eq.Ldef}).
Lower-centrality interactions agree with extrapolation from 
$p$-induced reactions.
\footnotesize } 
\end{center} 
\end{figure}

The NA50 Collaboration has studied the attenuation of $\jj$ 
by studying the ratio:

\begin{equation}
{{\cal A}_{\rm expt}}=
\frac{{\cal B}_{\mu\mu} \sigma(\jj)}{\sigma({\rm Drell-Yan})}\;\;.
\label{eq.exp}
\end{equation} as a function of the impact parameter~\cite{PLNA5099} .
Using Eq.~(\ref{eq.Ldef}), we can fit our Eq.~(\ref{eq.attenuation})
to the NA50 data, plotted as a function of $l$. 
The result is shown in Fig. \ref{isotherm}, for different values of the fireball temperature. 
A good fit can be achieved in the region of lower centrality, up to $l\simeq 5$~fm, 
thereby determining the fireball's temperature. Observe that peripheral interactions 
agree with extrapolation from $p$- or $d$-induced reactions (the points at $l\simeq 0$). 
From Fig.~\ref{isotherm} we find  $T\simeq 255$ MeV.

To be more precise on the temperature, however, we must consider the effect of 
nuclear absorption, due to the nucleon column density, 
which the $\jj$ has to traverse during the initial phase of the collision. 
Neglecting the  $\jj$ transverse momentum, 
the nuclear absorption factor is:
\begin{equation}
{{\cal A}_{\rm nucl.abs.}}=
{\rm exp}\left[-L\cdot\rho_{\rm nucl} \cdot \sigma_{\rm nucl}\right]
\label{eq.nuclatt}
\end{equation} 
$L$ is approximately equal to the longitudinal size of the 
overlap region of the colliding nuclei (Fig.~\ref{spheres}). 
To avoid unwanted distortions, we have interpolated the values of  
$L$ given by NA50 as functions of the impact parameter $b$ to 
obtain $L$ as a function of our length $l$ {\it via} 
Eq.~(\ref{eq.Ldef}).
%~\footnote{We acknowledge 
%a very informative discussion with L. Ramello on this point.}.  
A determination of the nuclear absorption cross section has been obtained 
by NA50 from the study of $\jj$ production in $p$--$A$ collisions. 
They find~\cite{sigmaN} :
\begin{equation}
\sigma_{\rm nucl}= 4.3 \pm 0.6~{\rm mb}
  \label{eq.sigman}
 \end{equation} This cross section is 
smaller than what would explain  the data 
in the low-centrality region with nuclear absorption 
only (this requires  $\sigma_{\rm nucl}\simeq 6$~mb~\cite{PLNA5099}) 
and points to 
the importance of dissociation induced by the comoving particles, which in 
fact turns out to dominate. 
In conclusion, we compare the experimental quantity 
Eq.~(\ref{eq.exp}) to the expression:

\begin{equation}
{{\cal A}_{\rm th}}(l)=C\cdot
{\rm exp}\left[-L(l)\cdot\rho_{\rm nucl} 
\cdot \sigma_{\rm nucl}\right]\cdot{\rm exp}\left[-6/10\cdot l
\langle\rho\cdot\sigma\rangle_T \right]\;\;,
\label{eq.totatt}
\end{equation} 
where $C$ is a normalization factor; 
$L$ and $l$ are obtained from the impact parameter 
$b$ following the NA50 prescription 
and Eq.~(\ref{eq.Ldef}), respectively.  
We use $\rho_{nucl} =0.17~ {\rm fm}^{-1}$,  $\sigma_{nucl}= 4.3$~mb.

\begin{figure}[ht] 
\begin{center} 
\epsfig{%bbllx=0.5cm,bblly=16cm,bburx=20cm,bbury=23cm, 
height=6.truecm, width=7.truecm, 
        figure=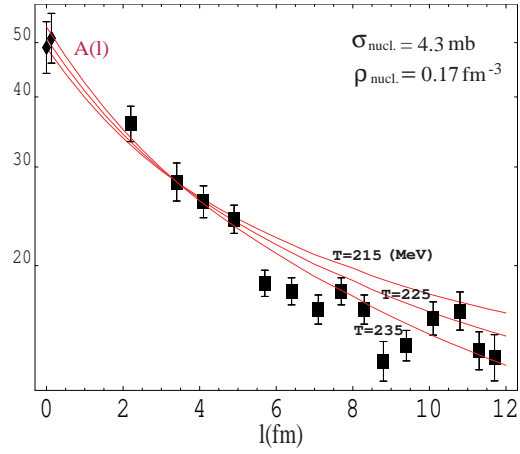}%\vskip1cm 
\caption{\label{withnabs} 
\footnotesize The attenuation function ${\cal A}$ 
given in Eq.~(\ref{eq.totatt}) 
as a function of the linear size of the fireball,
computed with $\rho_{\rm nucl}=0.17$~fm$^{-3}$
and $\sigma_{\rm nucl}=4.3$~mb. 
Lines correspond to a fixed temperature of the $\pi$ gas 
expressed in MeV. 
Experimental data: Pb-Pb collisions from 
NA50~\cite{PLNA5099}. Diamonds represent 
$p-(p,d)$~(NA51) data points~\cite{na51}.}
\end{center} 
\end{figure}

The results are shown in Fig.~\ref{withnabs}. 
The fireball temperature that gives the best fit to the low-centrality 
data, up to $l\simeq 5$~fm, is now $T= 225$~MeV, 
and the corresponding energy density  
is $\epsilon\simeq 0.32~{\rm GeV}\cdot {\rm fm}^{-3}$. We have 
studied how the theoretical errors in the coupling constants 
determined in~\cite{couplings} affect the determination of $T$. We find
$T=225\pm 15$~MeV.

Going towards higher centrality, we expect an 
increase in the energy density of the fireball. 
One effect that is easy to take into account 
is the increase in the surface density of the 
nucleons participating in the collision. 
The energy density of the fireball is, in fact, 
proportional to the factor~\cite{heavyreaction}: 
\begin{equation}
\frac{\rho_{\rm nucl}\cdot V(b)}{S(b)}= \frac{A}{S}\cdot g(b/R)
\end{equation}

\begin{equation}
g(b/R)=\frac{2}{3}\pi\frac{(1-b/2R)^2 (1+b/4R)}{{\rm arccos}(b/2R)- (b/2R) 
\sqrt{1-b^2/4 R^2}}\;\;,
\end{equation} with $R=A^{1/3}r_0$, the 
nucleon radius $r_0=1.1$~fm~\cite{abreu93}.

Starting from $l_0\simeq 5$~fm and $T_0=225$~MeV, 
the geometrical factor $g(2-l/R)$ is used 
to extrapolate the variation of temperature with $l$ according to:

\begin{equation}
T(l)=T_0 \left(\frac{g(2-l/R)}{g(2-l_0/R)}\right)^{1/4}\label{eq.modult}\;\;.
\end{equation} 
Once we know $T(l)$ we compute the corresponding attenuation 
function defined as: 
\begin{equation}
\tilde{{\cal A}}(l)=C\cdot
{\rm exp}\left[-L(l)\rho_{\rm nucl} 
\cdot \sigma_{\rm nucl}\right] {\rm exp}\left[- 6/10\cdot 
l\cdot \langle \rho \cdot \sigma
\rangle_{T(l)} \right]\label{eq.attenmod}\;\;.
\end{equation} 

\begin{figure}[ht] 
\begin{center} 
\epsfig{%bbllx=0.5cm,bblly=16cm,bburx=20cm,bbury=23cm, 
height=6.truecm, width=7.truecm, 
        figure=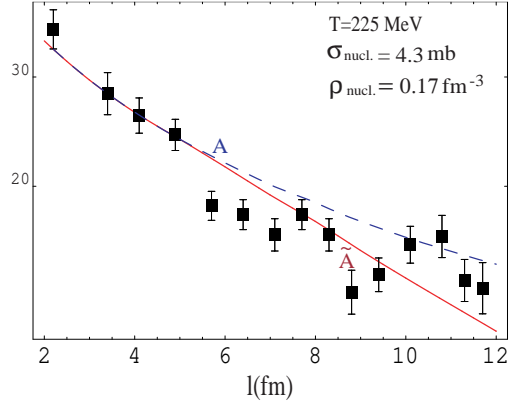}%\vskip1cm 
\caption{\label{thtotal} 
\footnotesize Complete calculation of
the attenuation function. $T=225$~MeV, $\rho_{\rm nucl}=0.17$~fm$^{-3}$
and $\sigma_{\rm nucl}=4.3$~mb. The 
solid line includes the effect of the 
increase of nucleon density per unit surface with decreasing 
impact parameter. The dashed line, given for reference,
is the same as in Fig.~6. Data from NA50 Pb-Pb 
collisions~\cite{PLNA5099}.
}
\end{center} 
\end{figure}

Fig~\ref{thtotal} shows the final best prediction of the 
conventional absorption effects due to the pion gas 
fireball and to the nuclear matter (solid curve). 

Fitting the data of low-centrality, 
we are able to determine with some accuracy the temperature 
of the fireball, assumed to be an ideal pion gas. 
The temperature is remarkably close to, perhaps somewhat higher than, the critical temperature computed in lattice calculations~\cite{Karsch}, ${T_{c}} \sim 180$~MeV. We recall that in SU(3) gauge theory, $T_c = 260$~MeV  while lattice results with $n_f =2$ suggest $T_c = 170$~MeV. If the same high temperature limit is assumed
for $n_f > 0$ and SU(3), then for $n_f= 2~{\rm or}~3$, the $\psi^\prime$ breaks at 
$190$--$200$~MeV and the $\jj$ itself would not break up 
until $T > 2 T_c$ \cite{vogt}.
Departures of the estimated temperature from the real one could be due 
either to an underestimation of the cross section, or to the inadequacy of the perfect 
gas approximation, or to both. Large deviations from the perfect gas 
are found in numerical simulations below the critical 
temperature~\cite{Karsch}. Further work to improve on the perfect Bose 
gas approximation and to obtain a better calibration of the crucial parameters of 
the fireball, temperature and energy density, is needed.

The energy density can be compared with the formula given in 
ref.~\cite{heavyreaction}:
\begin{equation}
\epsilon=\frac{A(b)}{S(b)}\left(\frac{dE}{dy}\right)\frac{1}{ct}\;\;,
\end{equation} $A(b)$ and $S(b)$ are the number of nucleons that participate 
in the collision and the overlap area, respectively, as functions of the impact 
parameter $b$; $dE/dy$ is the energy deposited per unit rapidity and $ct$, the 
longitudinal length of the fireball, is related to the time $t$ it takes for the 
nuclei to separate completely.
The energy deposited per unit rapidity is estimated as follows~\cite{heavyreaction}:
\begin{equation}
\frac{dE}{dy}=\frac{dN_{ch}}{dy}\left(1+
\frac{N_{\rm neutr}}{N_{ch}}\right)\langle E\rangle 
\simeq {3 \times 1.5 \times 400}~{\rm MeV} =1.8~{\rm GeV}\;\;.
\end{equation}

For central Pb-Pb collisions, one has:

\begin{equation}
\frac{A(b=0)}{S(b=0)}=\frac{A}{\pi R^2}=\frac{A^{1/3}}{\pi r_{0}^{2}}=
1.5~{\rm fm}^{-2}\;\;,
\end{equation} 

For Pb and $L=5$~fm
we find $g = 0.7$. 
In conclusion, for Pb-Pb collisions we find:

\begin{equation}
\epsilon=0.95~{\rm GeV}/{\rm fm}^{3} \cdot 
\left(\frac{2~{\rm fm}}{ct}\right)\,\,\,\,\,
(L=5~{\rm fm}) 
\end{equation}
\begin{equation}
\epsilon=1.35~{\rm GeV}/{\rm fm}^{3} \cdot \left(\frac{2~{\rm fm}}{ct}
\right)\,\,\,\,\, (L=2R=13~{\rm fm})\;\;.
\end{equation}

\section{Conclusions and Outlook}

Our final results are  displayed in Figs.~\ref{withnabs} 
and~\ref{thtotal}. 
In the lower-centrality region, where we expect to be still 
in the deconfined phase, we find $T=225\pm 15$~MeV 
corresponding to the energy density 
${\epsilon} \simeq 0.32~{\rm GeV}/{\rm fm}^{3}$. 
The value of the energy density is not inconsistent with the Bjorken formula. 
The temperature, however, is 
somewhat higher than the critical temperature computed in lattice 
calculations. 

The extrapolation to higher centrality keeps into account only the increase of 
the average nucleon number per unit area (an admittedly crude approximation). This 
leads to a downward bending of the $J/ \psi$ production as a function of $l$. However, 
the behaviour is too smooth to 
reproduce the rather sharp break shown by the NA50 data. 

Our analysis suggests that the simple pion gas description ceases to be valid 
at higher values of $l$, thus lending some support to the formation of a new phase. 
To confirm or disprove this, it 
would be crucial to analyse in the same way, i.e. as a function of $l$, the data 
obtained at the SPS and RHIC for other observables, such as strange particles production. In a pion 
gas, given the relatively low temperatures that we find in the low-centrality region, we 
would expect very few strange particles to be produced in the central region. This would not 
apply to the deconfined phase, where the Boltzmann suppression of strange 
quarks with $m_s = 150$~MeV would not be operative for temperatures 
around $225$~MeV.

\section*{Appendix}
We report the amplitudes corresponding to the tree--level 
diagrams for the $DD^*$ and $D^*D^*$ final states.
The diagrams and the corresponding amplitudes are as follows (dashed lines
for spinless particles, continuous lines for spin-1 particles); we use the
same notation as in~\cite{couplings}.
\begin{figure}[hbt]
\SetScale{0.8}
\noindent
{\begin{picture}(300,100)(0,0)
\DashLine(0,50)(120,50){5}
\Line(0,0)(60,0)
\DashLine(60,0)(120,0){5}
\Vertex(60,50){2}
\Vertex(60,0){2}
\Line(60,0)(60,50)
\put(0,50){$\pi$}
\put(0,-20){$J/\psi$}
\put(90,50){$D$}
\put(90,-20){$\bar D$}
\put(55,15){$D^*$}
\put(40,-35){$(1a)$}
\put(-12,0){$p_1$}
\put(-12,40){$p_2$}
\put(102,40){$p_3$}
\put(102,0){$p_4$}
\end{picture}
\hspace*{7em}
\begin{picture}(0,0)(220,0)
\DashLine(0,50)(60,50){5}
\DashLine(60,50)(120,0){5}
\Line(0,0)(60,0)
\DashLine(60,0)(120,50){5}
\Vertex(60,50){2}
\Vertex(60,0){2}
\Line(60,0)(60,50)
\put(0,50){$\pi$}
\put(0,-20){$J/\psi$}
\put(90,50){$D$}
\put(90,-20){$\bar D$}
\put(30,15){$D^*$}
\put(40,-35){$(1b)$}
\put(-12,0){$p_1$}
\put(-12,40){$p_2$}
\put(102,40){$p_3$}
\put(102,0){$p_4$}
\end{picture}
\hspace*{7em}
\begin{picture}(0,0)(130,0)
\DashLine(0,50)(120,0){5}
\Line(0,0)(60,25)
\DashLine(60,25)(120,50){5}
\Vertex(60,25){2}
\put(0,50){$\pi$}
\put(0,-20){$J/\psi$}
\put(90,50){$D$}
\put(90,-20){$\bar D$}
\put(40,-35){$(1c)$}
\put(-12,0){$p_1$}
\put(-12,40){$p_2$}
\put(102,40){$p_3$}
\put(102,0){$p_4$}
\end{picture}
}
\end{figure}

\begin{figure}[hbt]
\SetScale{0.8}
\noindent
{\begin{picture}(230,100)(0,0)
\DashLine(0,50)(60,50){5}
\Line(60,50)(120,50)
\Line(0,0)(60,0)
\DashLine(60,0)(120,0){5}
\Vertex(60,50){2}
\Vertex(60,0){2}
\DashLine(60,0)(60,50){5}
\put(0,50){$\pi$}
\put(0,-20){$J/\psi$}
\put(90,50){$D^*$}
\put(90,-20){$\bar D$}
\put(55,15){$D$}
\put(40,-35){$(2a)$}
\put(-12,0){$p_1$}
\put(-12,40){$p_2$}
\put(102,40){$p_3$}
\put(102,0){$p_4$}
\end{picture}
\hspace*{7em}
\begin{picture}(0,0)(140,0)
\DashLine(0,50)(60,50){5}
\Line(60,50)(120,50)
\Line(0,0)(60,0)
\DashLine(60,0)(120,0){5}
\Vertex(60,50){2}
\Vertex(60,0){2}
\Line(60,0)(60,50)
\put(0,50){$\pi$}
\put(0,-20){$J/\psi$}
\put(90,50){$D^*$}
\put(90,-20){$\bar D$}
\put(55,15){$D^*$}
\put(40,-35){$(2b)$}
\put(-12,0){$p_1$}
\put(-12,40){$p_2$}
\put(102,40){$p_3$}
\put(102,0){$p_4$}
\end{picture}
\hspace*{-8.3em}
\vspace*{7em}
\begin{picture}(0,100)(230,100)
\DashLine(0,50)(60,50){5}
\DashLine(60,50)(120,0){5}
\Line(0,0)(60,0)
\Line(60,0)(120,50)
\Vertex(60,50){2}
\Vertex(60,0){2}
\Line(60,0)(60,50)
\put(0,50){$\pi$}
\put(0,-20){$J/\psi$}
\put(90,50){$D^*$}
\put(90,-20){$\bar D$}
\put(30,15){$D^*$}
\put(40,-35){$(2c)$}
\put(-12,0){$p_1$}
\put(-12,40){$p_2$}
\put(102,40){$p_3$}
\put(102,0){$p_4$}
\end{picture}
\hspace*{7em}
\vspace*{7em}
\begin{picture}(0,100)(140,100)
\DashLine(0,50)(120,0){5}
\Line(0,0)(60,25)
\Line(60,25)(120,50)
\Vertex(60,25){2}
\put(0,50){$\pi$}
\put(0,-20){$J/\psi$}
\put(90,50){$D^*$}
\put(90,-20){$\bar D$}
\put(40,-35){$(2d)$}
\put(-12,0){$p_1$}
\put(-12,40){$p_2$}
\put(102,40){$p_3$}
\put(102,0){$p_4$}
\end{picture}
}
\end{figure}

\begin{figure}[hbt]
\SetScale{0.8}
\noindent
{\begin{picture}(360,100)(0,0)
\DashLine(0,50)(60,50){5}
\Line(60,50)(120,50)
\Line(0,0)(60,0)
\Line(60,0)(120,0)
\Vertex(60,50){2}
\Vertex(60,0){2}
\DashLine(60,0)(60,50){5}
\put(0,50){$\pi$}
\put(0,-20){$J/\psi$}
\put(90,50){$D^*$}
\put(90,-20){$\bar D^*$}
\put(55,15){$D$}
\put(40,-35){$(3a)$}
\put(-12,0){$p_1$}
\put(-12,40){$p_2$}
\put(102,40){$p_3$}
\put(102,0){$p_4$}
\end{picture}
\hspace*{7em}
\begin{picture}(0,0)(280,0)
\DashLine(0,50)(60,50){5}
\Line(60,50)(120,0)
\Line(0,0)(60,0)
\Line(60,0)(120,50)
\Vertex(60,50){2}
\Vertex(60,0){2}
\DashLine(60,0)(60,50){5}
\put(0,50){$\pi$}
\put(0,-20){$J/\psi$}
\put(90,50){$D^*$}
\put(90,-20){$\bar D^*$}
\put(30,15){$D$}
\put(40,-35){$(3b)$}
\put(-12,0){$p_1$}
\put(-12,40){$p_2$}
\put(102,40){$p_3$}
\put(102,0){$p_4$}
\end{picture}
\hspace*{7em}
\begin{picture}(0,0)(200,0)
\DashLine(0,50)(60,50){5}
\Line(60,50)(120,50)
\Line(0,0)(60,0)
\Line(60,0)(120,0)
\Vertex(60,50){2}
\Vertex(60,0){2}
\Line(60,0)(60,50)
\put(0,50){$\pi$}
\put(0,-20){$J/\psi$}
\put(90,50){$D^*$}
\put(90,-20){$\bar D^*$}
\put(55,15){$D^*$}
\put(40,-35){$(3c)$}
\put(-12,0){$p_1$}
\put(-12,40){$p_2$}
\put(102,40){$p_3$}
\put(102,0){$p_4$}
\end{picture}
\hspace*{-16.3em}
\vspace*{7em}
\begin{picture}(0,100)(280,100)
\DashLine(0,50)(60,50){5}
\Line(60,50)(120,0)
\Line(0,0)(60,0)
\Line(60,0)(120,50)
\Vertex(60,50){2}
\Vertex(60,0){2}
\Line(60,0)(60,50)
\put(0,50){$\pi$}
\put(0,-20){$J/\psi$}
\put(90,50){$D^*$}
\put(90,-20){$\bar D^*$}
\put(30,15){$D^*$}
\put(40,-35){$(3d)$}
\put(-12,0){$p_1$}
\put(-12,40){$p_2$}
\put(102,40){$p_3$}
\put(102,0){$p_4$}
\end{picture}
\hspace*{7em}
\vspace*{7em}
\begin{picture}(0,100)(200,100)
\DashLine(0,50)(60,25){5}
\Line(60,25)(120,0)
\Line(0,0)(60,25)
\Line(60,25)(120,50)
\Vertex(60,25){2}
\put(0,50){$\pi$}
\put(0,-20){$J/\psi$}
\put(90,50){$D^*$}
\put(90,-20){$\bar D^*$}
\put(40,-35){$(3e)$}
\put(-12,0){$p_1$}
\put(-12,40){$p_2$}
\put(102,40){$p_3$}
\put(102,0){$p_4$}
\end{picture}
}
\end{figure}

In the case in which the final state is $DD^*$:
\ba
A_{2a}&=&-\frac{2g_{\pi DD^*}g_{\jj DD}}{t-M_D^2}(p_2\cdot \epsilon(p_3))(p_4\cdot
\epsilon(p_1))\nonumber\\
A_{2b}&=& i\frac{g_{\pi D^*D^*}g_{\jj DD}}{M_D(t-M_{D^*}^2)}
\epsilon(\epsilon(p_1),p_4,p_1,\rho)\times\nonumber\\
&& \epsilon(\epsilon(p_3),p_2,p_3,\nu) \Pi^{\nu\rho}(p_3-p_2)\nonumber\\
A_{2c}&=&\frac{g_{\pi DD}g_{\jj D^* D^*} }{u-M_{D^*}^2}p_{2\rho} \left(
(-(p_3\cdot\epsilon(p_1))\epsilon_\mu(p_3)+ \right.\nonumber\\
&&\left.
((p_3-p_1)\cdot\epsilon(p_1))\epsilon_\mu(p_3)+
(\epsilon(p_1)\cdot\epsilon(p_3))p_{3\mu})\times 
\right.\nonumber\\&&\left.
\Pi^{\mu\rho}(p_3-p_1,M_{D^*}^2)+((p_3-p_1)\cdot\epsilon(p_3))\epsilon_\nu(p_1)
\times\right.\nonumber\\&&\left.
\Pi^{\nu\rho}(p_3-p_1,M_{D^*}^2)\right)\nonumber\\
A_{2d}&=&-p_2^\nu \epsilon^\mu(p_1)\left( g_3 M_D^{-2} (p_{3\nu}
p_{4\mu} (p_4\cdot\epsilon(p_3))- p_{3\mu} p_{4\nu} (p_4\cdot\epsilon(p_3)))
-\right.\nonumber\\  &&\left.
g_2(p_{4\nu}\epsilon_{\mu}(p_3)+p_{3\nu}\epsilon_{\mu}(p_3))+g_1(g_{\mu\nu}
(p_4\cdot\epsilon(p_3))  
+\right.\nonumber\\  &&\left.
2 p_{3\mu} \epsilon_{\nu}(p_3))  \right)\nonumber
\ea

The amplitudes for the $D^*D^*$ final state are:~\ba
A_{3a}&=& i\frac{g_{\pi DD^*}g_{\jj DD^*}}{t-M_D^2} \epsilon(\epsilon(p_1),
\epsilon(p_4),p_3-p_2,p_4)(p_2\cdot\epsilon(p_3))\nonumber\\
A_{3b}&=&i\frac{g_{\pi D^*D^*}g_{\jj DD^*}}{u-M_D^2} \epsilon(\epsilon(p_1),
p_3-p_2,\epsilon(p_3),p_3)(p_2\cdot\epsilon(p_4))\nonumber\\
A_{3c}&=&\frac{g_{\pi D^*D^*} g_{\jj D^* D^*}}{t-M_{D^*}^2} (
\epsilon(p_3-p_2,\tau,p_3,\epsilon(p_3))\times
\nonumber\\&&
\Pi^{\lambda\tau}(p_3-p_2,M_{D^*})
(p_{4\lambda}(\epsilon(p_1)\cdot\epsilon(p_4))-
\nonumber\\&&
\epsilon_\lambda(p_1)
((p_3-p_2)\cdot\epsilon(p_4))-\epsilon_\lambda(p_4)(p_4\cdot\epsilon(p_1))+
\nonumber\\&&
\epsilon_\lambda(p_4)((p_3-p_2)\cdot\epsilon(p_1)))\nonumber\\
A_{3d}&=&\frac{g_{\pi D^*D^*}g_{\jj D^*D^*}}{u-M_{D^*}^2}
\epsilon(p_4,\epsilon(p_4),p_3-p_1,\beta)
\times\nonumber\\&&
\Pi^{\beta\lambda}(p_3-p_1,M_{D^*})
(-\epsilon(p_3)\cdot\epsilon(p_1)p_{3\lambda}+
(p_3\cdot\epsilon(p_1))\epsilon_\lambda(p_3)  +
\nonumber\\&&
((p_3-p_1)\cdot\epsilon(p_1))\epsilon_\lambda(p_3)-
\nonumber\\&&
((p_3-p_1)\cdot\epsilon(p_3))\epsilon_\lambda(p_1))\nonumber\\
A_{3e}&=&iM_D p_{2\nu}\epsilon_{\mu}(p_1)\left(
\frac{g_8}{M_D^2}\epsilon(\mu,\epsilon(p_3),\epsilon(p_4),p_3)p_3^\nu-
%\right.\nonumber\\&&\left.
\frac{g_6}{M_D^2}\epsilon(\mu,\epsilon(p_3),\epsilon(p_4),p_3)p_4^\nu+
\right.\nonumber\\&&\left.
\frac{g_1+2g_7}{M_D^2}\epsilon(\mu,p_4,\epsilon(p_4),p_3)\epsilon^\nu(p_3)+
%\right.\nonumber\\&&\left.
\frac{g_9}{M_D^2}\epsilon(\mu,\epsilon(p_3),p_4,\epsilon(p_4))p_3^\nu+
\right.\nonumber\\&&\left.
\frac{g_1}{M_D^2}\epsilon(\nu,p_4,\epsilon(p_3),\epsilon(p_4))p_4^\mu+
%\right.\nonumber\\&&\left.
\frac{g_5}{M_D^2}\epsilon(\mu,p_4,\epsilon(p_3),\epsilon(p_4))p_4^\nu+
\right.\nonumber\\&&\left.
\frac{g_1+g_7}{M_D^2}\epsilon(\nu,p_4,\epsilon(p_3),p_3)\epsilon^\mu(p_4)-
%\right.\nonumber\\&&\left.
g_4\epsilon(\mu,\nu,\epsilon(p_3),\epsilon(p_4))-
\right.\nonumber\\&&\left.
\frac{g_5+g_7}{M_D^4}\epsilon(p_4,\epsilon(p_3),\epsilon(p_4),p_3)
p_3^\mu p_3^\nu +
%\right.\nonumber\\&&\left.
\frac{g_9}{M_D^4}\epsilon(p_4,\epsilon(p_3),\epsilon(p_4),p_3)
p_4^\mu p_4^\nu-
\right.\nonumber\\&&\left.
\frac{g_7}{M_D^2}( 2\epsilon(\nu,\epsilon(p_3),\epsilon(p_4),p_3)p_3^\mu-
%\right.\nonumber\\&&\left.
\epsilon(p_4,\nu,\epsilon(p_4),p_3)\epsilon^\mu(p_3)+
\right.\nonumber\\&&\left.
2\epsilon(\mu,p_4,\epsilon(p_3),p_3)\epsilon^\nu(p_4)-
%\right.\nonumber\\&&\left.
g^{\mu\nu}\epsilon(p_4,\epsilon(p_3),\epsilon(p_4),p_3))
\right)
\nonumber
\ea
\newpage
\acknowledgements{We would like to thank Paolo Giubellino and Luciano Ramello
for useful information on NA50 data and Michelangelo Mangano for interesting
discussions.

ADP thanks K.~Eskola, R.~Gatto, G.~Nardulli for 
informative discussions and A. Strumia for a useful Mathematica Macro. 
FP acknowledges some remarks by G. Zambotti. 

VR would like to acknowledge financial support from CONACYT, M\'exico.}


\begin{thebibliography}{99}

\bibitem{couplings}  A. Deandrea, G. Nardulli and A.D. Polosa,
Phys.\ Rev.\ {\bf D68} (2003) 034002; see also M. Bedjidian {\it et al.},
``Hard probes in heavy ion collisions at the LHC: heavy flavor physics'',
arXiv:hep-ph/0311048.

\bibitem{proto} A. Deandrea, N. Di Bartolomeo, R. Gatto, G. Nardulli and A.D. Polosa,
Phys.\ Rev.\ {\bf D58} (1998) 034004; see also A.D.~Polosa,
``The CQM Model'', Riv. Nuovo Cim. Vol. {\bf 23}, N. 11 (2000).

\bibitem{cina}
Z.W.~Lin and C.M.~Ko,
% ``A model for $J/\psi$ absorption in
%hadronic matter,''
Phys.\ Rev.\ {\bf C62} (2000) 034903, arXiv:nucl-th/9912046;
K.~L.~Haglin and C.~Gale,
%``Hadronic interactions of the J/psi,''
Phys.\ Rev. {\bf C63} (2001) 065201, arXiv:nucl-th/0010017;
Y.~Oh, T.~Song and S.H.~Lee,
%``$J/\psi$ absorption by pi and rho mesons in meson exchange model
%with anomalous parity interactions,''
Phys.\ Rev.\ {\bf C63} (2001) 034901, arXiv:nucl-th/0010064.
\bibitem{exp}
S.~Ahmed {\it et al.}  [CLEO Collaboration],
%``First measurement of Gamma(D*+),''
Phys.\ Rev.\ Lett.\  {\bf 87} (2001) 251801, arXiv:hep-ex/0108013.
\bibitem{qcdsr}
R.~D.~Matheus, F.~S.~Navarra, M.~Nielsen and R.~Rodrigues da Silva,
%``The J/psi D D vertex in QCD sum rules,''
Phys.\ Lett.\ {\bf B541} (2002) 265, arXiv:hep-ph/0206198
F.~O.~Duraes, S.~H.~Lee, F.~S.~Navarra and M.~Nielsen,
%``J/psi dissociation by pions in QCD,''
arXiv:nucl-th/0210075; for a nice review: R.~Rapp and L.~Grandchamp, 
arXiv:hep-ph/0305143.

\bibitem{sigmaN} B.~Alessandro {\it et al.} (NA50 Coll.), Nucl.\ Phys.\ 
{\bf A715} (2003) 679c. 

\bibitem{heavyreaction}  J. D. Bjorken, Phys.\ Rev.\ {\bf D27} (1983) 140.

\bibitem{PLNA5099} M. C. Abreu {\it et al.} (NA50 Coll.), 
Phys.\ Lett.\ {\bf B450} (1999) 456.

\bibitem{Karsch}  F. Karsch, Univ. of Bielefeld preprint BI-TP 2004/04, 
arXiv: hep-1at/0401031.

\bibitem{rapp} F.~O.~Duraes {\it et al.}  Phys.\ Rev.\  {\bf C68} (2003) 035208, arXiv:nucl-th/0211092; 
R.~Rapp and L.~Grandchamp, J. \ Phys. \  
{\bf  G30} (2004) 5305, arXiv:hep-ph/0305143.

\bibitem{abreu93} M.~C.~ Abreu {\it et al.}, Phys. Lett. {\bf B410} (1993) 337.

\bibitem{vogt} R. Vogt, Phys. \ Rept. {\bf 310} (1999) 197.

\bibitem{na51} M. C.~Abreu {\it et al.}, (NA51 Coll.)
Phys.\ Lett.\ {\bf B438} (1998) 35.

\end{thebibliography}
\end{document}